\documentclass[prl,twocolumn,letterpaper,showpacs,groupedaddress,superscriptaddress,nofootinbib,floatfix,preprintnumbers,tightenlines]{revtex4-1}
\usepackage{hyperref,amssymb,amsmath,graphicx,xcolor}
\usepackage{bm}
\usepackage{mathtools}

\def\si{^1 \hskip -0.03in S _0}

\begin{document} 
\begin{figure}[!t]
\vskip -1.1cm
\leftline{
\includegraphics[width=3.0 cm]{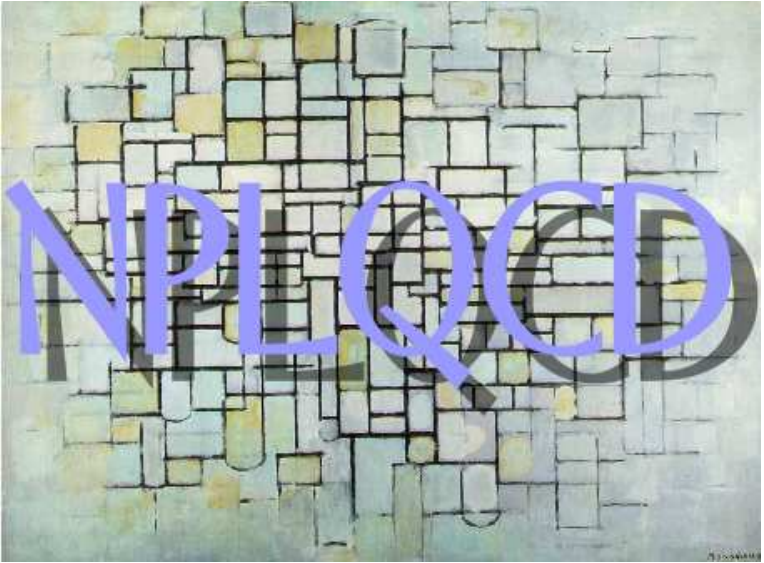}}
\vskip -0.5cm
\end{figure}

\title{Taming the Signal-to-Noise Problem in Lattice QCD by Phase Reweighting}

\author{Michael L. Wagman} 
\affiliation{Institute for Nuclear Theory, University of Washington, Seattle, WA 98195-1550, USA}
\affiliation{Department of Physics,
	University of Washington, Box 351560, Seattle, WA 98195, USA}

\author{Martin J. Savage}
\affiliation{Institute for Nuclear Theory, University of Washington, Seattle, WA 98195-1550, USA}
 \affiliation{Department of Physics,
	University of Washington, Box 351560, Seattle, WA 98195, USA}

\collaboration{NPLQCD Collaboration}

\date{\today}

\preprint{INT-PUB-17-012}


\begin{abstract}
Path integrals describing quantum many-body systems can be calculated with Monte Carlo sampling techniques, 
but average quantities are often subject to signal-to-noise ratios that degrade exponentially with time.
A phase-reweighting technique inspired by recent observations of random walk statistics in correlation functions
is proposed that allows energy levels to be extracted from late-time correlation functions with time-independent signal-to-noise ratios.
Phase reweighting effectively includes dynamical refinement of source magnitudes
but introduces a bias
associated with the phase.
This bias can be removed
by performing 
an extrapolation,
but at the expense of
re-introducing a signal-to-noise problem.
Lattice Quantum Chromodynamics calculations 
of the $\rho^+$ and nucleon masses and of the $\Xi\Xi(\si)$ binding energy
show consistency between standard results obtained using earlier-time correlation functions and
phase-reweighted results using late-time correlation functions inaccessible to standard statistical analysis methods.
\end{abstract}

\maketitle


The signal-to-noise (StN) problem inherent to Monte Carlo sampling of quantum mechanical correlation functions
provides a substantial impediment to precision calculations of multi-particle systems across many areas of physics, 
from Lattice Quantum Chromodynamics (LQCD) and nuclear many-body calculations to calculations of the properties of materials.
In LQCD,
where the quantum fields responsible for the strong and electromagnetic forces are
sampled numerically on a discretized spacetime to calculate path integrals,
the StN problem has restricted calculations to mesons, the nucleon, and the lightest few nuclei.
Ideally, calculations of larger nuclei and of the dense matter present in 
the interior of neutron stars 
would also be performed directly with LQCD, but the StN problem provides a  substantial roadblock.

StN problems in LQCD have been studied since the pioneering works of
Parisi~\cite{Parisi:1983ae} and Lepage~\cite{Lepage:1989hd}, 
and arises when there are states contributing to a variance correlation function with less than
twice the energy of the ground state of the correlation function.
Correlation functions describing one or more baryons in LQCD have exponentially degrading StN ratios at late (Euclidean) times, 
with the  argument of the exponent increasing with the number of baryons~\cite{Beane:2010em}.
The statistical distributions of correlation functions sampled in Monte Carlo calculations have interesting 
features~\cite{Beane:2009kya,Beane:2009gs,Beane:2010em,Endres:2011jm,Endres:2011er,Endres:2011mm,Lee:2011sm,DeGrand:2012ik,Grabowska:2012ik,Nicholson:2012xt,Beane:2014oea}, 
and in particular the logarithms of LQCD correlation functions exhibit characteristics of L{\'e}vy Flights associated with 
heavy-tailed Stable Distributions~\cite{Wagman:2016bam}.
At early and intermediate times, the distribution of the real parts of nucleon correlation functions is asymmetric 
with odd moments that fall exponentially with the nucleon mass, $M_N$,
and, in contrast, even moments that
fall exponentially with the pion mass, $M_\pi$~\cite{Beane:2014oea}.  
This leads to a distribution at late times
that is symmetric and non-Gaussian and a 
nucleon StN ratio proportional to $\sim e^{-(M_N-3 M_\pi/2) t}$.
Sink optimization for baryon and multi-baryon systems~\cite{Beane:2009kya,Beane:2009gs,Beane:2010em,Beane:2012vq,Beane:2013br,Beane:2014oea,Detmold:2014rfa,Detmold:2014hla},
and more sophisticated variational methods in the mesonic sector~\cite{Michael:1985ne,Luscher:1990ck,Dudek:2007wv,Blossier:2009kd},
can extend the plateau region where correlation functions achieve approximate ground-state saturation to earlier times.
In this ``golden window,'' variance correlation functions have not yet achieved ground-state saturation
and StN degradation is exponentially less severe than at later times~\cite{Beane:2009kya,Beane:2009gs,Beane:2010em}.
At very late times, nucleon correlation functions enter a noise region
where standard statistical estimators, 
including the sample mean, 
become unreliable because of finite sample size effects associated with circular statistics~\cite{Wagman:2016bam}.

To begin extracting meaningful results from the noise region,
it is helpful to separately consider the magnitude and phase of nucleon correlation functions~\cite{Wagman:2016bam}.
The average nucleon magnitude is observed to be proportional to $\sim e^{-3M_\pi t/2}$ at late times
and does not exhibit a StN problem. 
In contrast, 
the average nucleon phase is observed to be proportional to $\sim e^{-(M_N-3 M_\pi/2) t}$ at late times and has a severe StN problem.
From this behavior, the StN problem in nucleon correlation functions was identified as a sign problem~\cite{Wagman:2016bam}.
The sign problem encountered in estimating the phase of a correlation function 
is spacetime extensive
and can be mitigated by restricting the time interval, $\Delta t$, over which the system contains specific conserved charges prior to measurement.
This restriction neglects correlations across distances larger than $\Delta t$
and creates a bias in ground-state energies that decreases exponentially with increasing $\Delta t$.

\begin{figure}[!t]
	\includegraphics[width=0.95 \columnwidth]{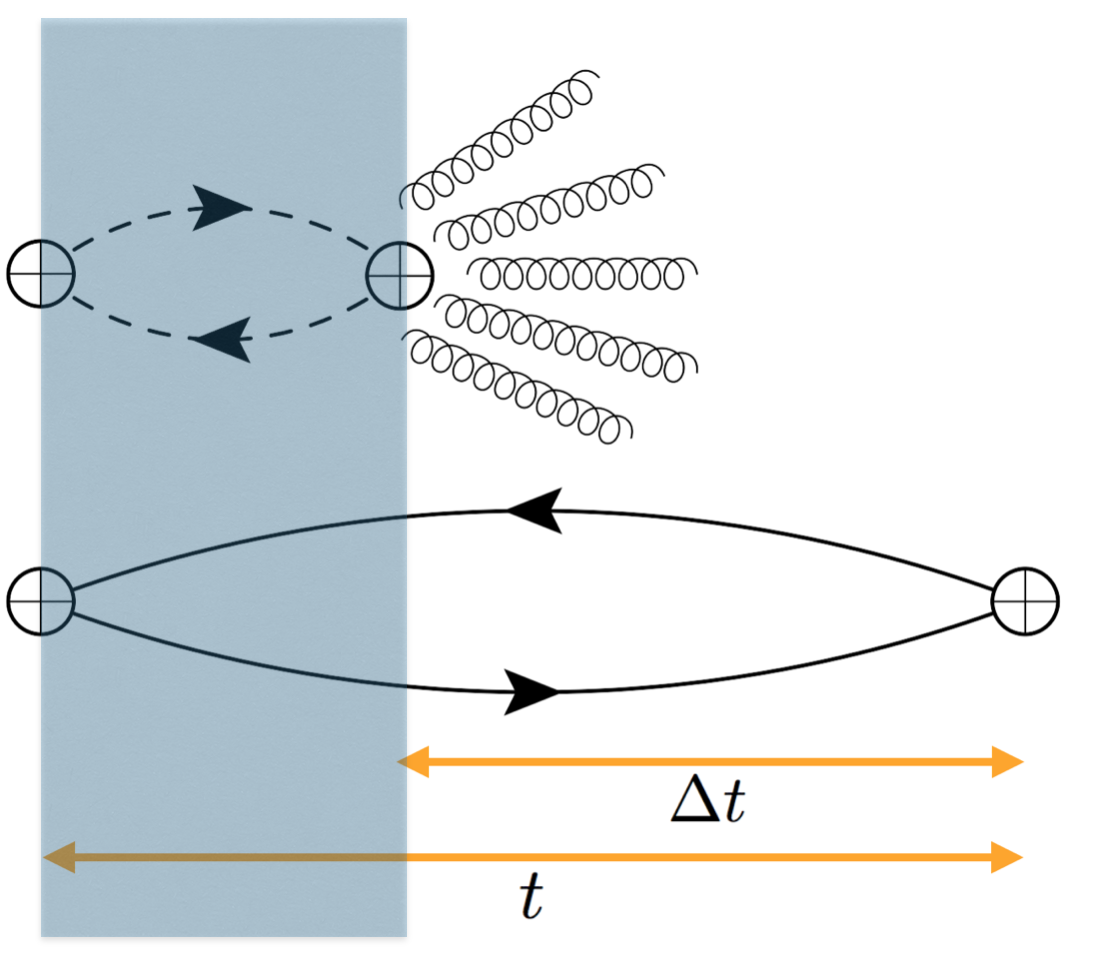}
		\caption{
	\label{fig:PRWcorrs} 
  The $\rho^+$-meson phase-reweighted correlation function $G^\theta_\rho(t,\Delta t)$ is a product
  of quark propagators forming $C_i^\rho(t)$, shown as solid lines,
  and a phase factor $e^{-i\theta_i^\rho(t-\Delta t)}$, shown as dashed propagator lines with reversed quark-charge arrows.
        Gluon lines indicate that phase reweighting introduces correlations
        associated with excitations produced at $t-\Delta t$
        and lead to bias when $\Delta t \neq t$.
        For momentum-projected correlation functions, 
        excitations involving correlated interactions between 
        $C_i^\rho(t)$ and $e^{-i\theta_i^\rho(t-\Delta t)}$ are suppressed by the spatial volume.
        $G^\theta_\rho(t,\Delta t)$ effectively includes a non-local source
        whose magnitude is dynamically refined for $t - \Delta t$ steps
        while the phase is held fixed (shaded region)
        before the full system is evolved for the last $\Delta t$ steps of propagation.
	}		
\end{figure}

This letter introduces a phase reweighting technique for LQCD correlation functions that allows ground-state energies to be extracted at
late times with StN constant in $t$.
By restricting the region where the complex phase associated with baryon number is allowed to evolve, 
phase reweighting makes $\Delta t$ independent of $t$,
but leads to a bias that must be systematically removed through extrapolation.
The StN problem re-emerges as exponential loss of precision with increasing $\Delta t$.

Analogous techniques are used in applications of Green's Function Monte Carlo (GFMC) methods to  
nuclear many-body systems where the phase 
of the wavefunction is held fixed until the system is close to its ground state, at which 
point the phase is released for final evolution~\cite{Zhang:1995zz,Zhang:1996us,Wiringa:2000gb,Carlson:2014vla}. 
Similar techniques are also used in Lattice Effective Field Theory (LEFT) calculations in which a Wigner-symmetric Hamiltonian, 
emerging from the large-N$_c$ limit of QCD~\cite{Kaplan:1995yg}, 
is used for initial time evolution before asymmetric perturbations  are added that introduce a sign problem~\cite{Lahde:2015ona}.
Phase reweighting shares physical similarities, and possibly formal connections, 
to the approximate factorization of domain-decomposed quark propagators recently suggested and explored by 
C$\grave{e}$, Giusti and Schaefer~\cite{Ce:2016idq,Ce:2016ajy,Ce:2016qto}.

LQCD calculations involve ensembles of a large number, $N$, of correlation functions $C_i(t)$, each calculated from a 
source on a particular gauge field configuration.
Expectation values $G(t)=\langle C_i(t) \rangle$ can be computed from sample averages $G(t) = {1\over N}\sum_i C_i(t)$ 
across field configurations importance sampled from the QCD vacuum probability distribution.
The ground-state energy of correlation functions can be accurately determined
from the late-time behavior of $G(t)$, but for generic correlation functions the StN problem
restricts the extraction of precise ground-state energy measurements to early and intermediate times.

Phase reweighted correlation functions are defined by
\begin{eqnarray}
G^{\theta}(t,\Delta t) & = & 
\langle e^{-i \theta_i (t-\Delta t)} \ C_i(t)  \rangle
\  \ ,
\label{eq:PRWdef}
\end{eqnarray}
where $\theta_i (t-\Delta t) = \text{arg}[ C_i(t-\Delta t) ]$.
Phase reweighting resembles limiting the approximate L{\'e}vy Flight of the correlation function phase 
to $\Delta t$ steps at late times, suggesting that $G^\theta(t,\Delta t)$ has a StN ratio that decreases exponentially with 
$\Delta t$ but is constant in $t$.
In the limit that $\Delta t \rightarrow t$, the reweighting factor approaches unity and $G^{\theta}(t,t) = G(t)$.
The exact correspondence $G^\theta(t,t)=G(t)$ 
gives phase reweighting
an advantage over our previously suggested estimator~\cite{Wagman:2016bam}  involving 
multiplication by $C_i^{-1}(t-\Delta t)$ rather than $e^{-i\theta_i(t-\Delta t)}$.
Phase reweighting also leads to more precise ground-state energy extractions than estimators involving reweighting with $C_i^{-1}(t-\Delta t)$;
multiplication by the heavy-tailed variable $|C_i(t-\Delta t)^{-1}|$ leads to increases variance.

Dynamical correlations between $C_i(t)$ and $e^{-i\theta_i(t-\Delta t)}$
lead to differences in  ground-state energies extracted from $G^\theta(t,\Delta t)$ and $G(t)$ for $t\neq \Delta t$. 
Locality suggests that these correlations should decrease exponentially with increasing $\Delta t$
at a rate controlled by the longest correlation length in the theory.
At asymptotically large $\Delta t$,
one-pion-exchange correlations are expected to provide the largest contributions to the bias.
These contributions will be
suppressed by factors involving the spatial volume in products of a momentum-projected correlation function with a momentum-projected phase factor.
Excitations involving the $\sigma$ meson, correlated two-pion exchange, and other light
excitations that do not change the quantum numbers of the system
are not volume-suppressed and
may dominate at small $\Delta t$.
Near-threshold bound states may have complicated small $\Delta t$ bias that is sensitive to the size of the spatial volume.

\begin{figure}[!t]
	\includegraphics[width=0.99 \columnwidth]{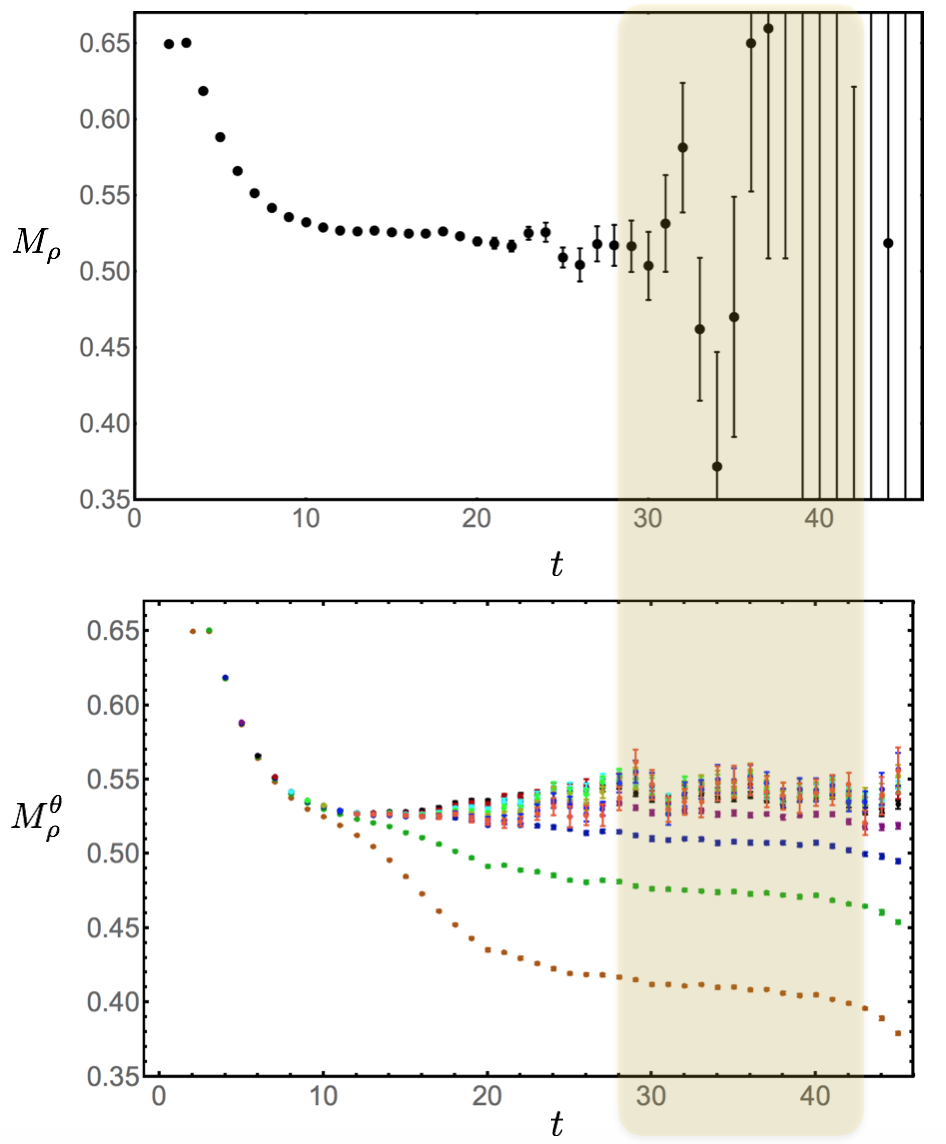}
		\caption{
	\label{fig:emps} 
  The upper panel shows the $\rho^+$ effective mass from the LQCD ensemble of Ref.~\cite{Orginos:2015aya}.
	The lower panel shows $M_\rho^\theta(t,\Delta t)$ with a range of fixed $\Delta t$'s.
        Temporal structure at later times arises from proximity to the midpoint of the lattice at $t=48$.
        The highlighted interval $t=28\rightarrow 43$ is used for correlated $\chi^2$ minimization fits of $M_\rho^\theta$.
	Masses and times are given in lattice units.
	}		
\end{figure}

The construction of $G^\theta$ is generic for any correlation function,
and is schematically depicted for the $\rho^+$ meson in Fig.~\ref{fig:PRWcorrs}.
In the plateau region of the  $\rho^+$ correlation function, the average of the  magnitude is approximately proportional to $e^{-M_\pi t}$, 
while the average of the phase factor\footnote{
The phases of isovector meson correlation functions are restricted to be discrete values $\theta_\rho = 0,\ \pi$ 
when interpolating operators in a Cartesian spin basis are used.  In forthcoming work, we demonstrate that circular statistics 
applies to real but non-positive isovector meson correlation functions.
} is approximately proportional to  $e^{-(M_\rho-M_\pi) t}$.
$G^\theta(t,\Delta t)$ is a product of these~two averages plus corrections arising from correlations between 
$C_i(t)$ and $e^{-i\theta_i(t-\Delta t)}$, and so at large $t$ and $\Delta t$ it is expected to have the form 
\begin{eqnarray}
G^{\theta}(t,\Delta t) & \sim &
e^{-M_\pi (t-\Delta t) } e^{-M_\rho \Delta t} \left(\alpha+\beta e^{-\delta M_{\rho} \Delta t } +  ... \right),
\;\;\;\;
\label{eq:PRWtdep}
\end{eqnarray}
where $M_\rho +\delta M_{\rho}$ is the energy of the lowest-lying excited state of the $\rho^+$
leading to appreciable correlations between $C_i(t)$ and $e^{-i\theta_i(t-\Delta t)}$,
and $\alpha$ and $\beta$ are overlap factors that cannot be determined with general arguments but can be calculated with LQCD.
The ellipses denote further-suppressed contributions from higher-lying states.
A phase-reweighted effective mass can be defined as
$M^\theta=\log\left( G^{\theta}(t,\Delta t) /G^{\theta}(t+1,\Delta t+1) \right)$, which reduces to the standard effective mass definition when $\Delta t\rightarrow t$.
For the $\rho^+$ meson, the form of the correlation function given in Eq.~(\ref{eq:PRWtdep}) leads to
\begin{eqnarray}
M_\rho^\theta(t, \Delta t) & = & M_\rho \ +\  c\ \delta M_{\rho} e^{-\delta M_{\rho} \Delta t}\ +\ ...
\ \ ,
\label{eq:PRWEMP}
\end{eqnarray}
at large $t$,
where $c=\beta/\alpha$ and the ellipses denote higher order contributions which are exponentially suppressed with
$\Delta t$ and standard excited state contributions that are exponentially suppressed with $t$.

\begin{figure}[!t]
	\includegraphics[width=0.99 \columnwidth]{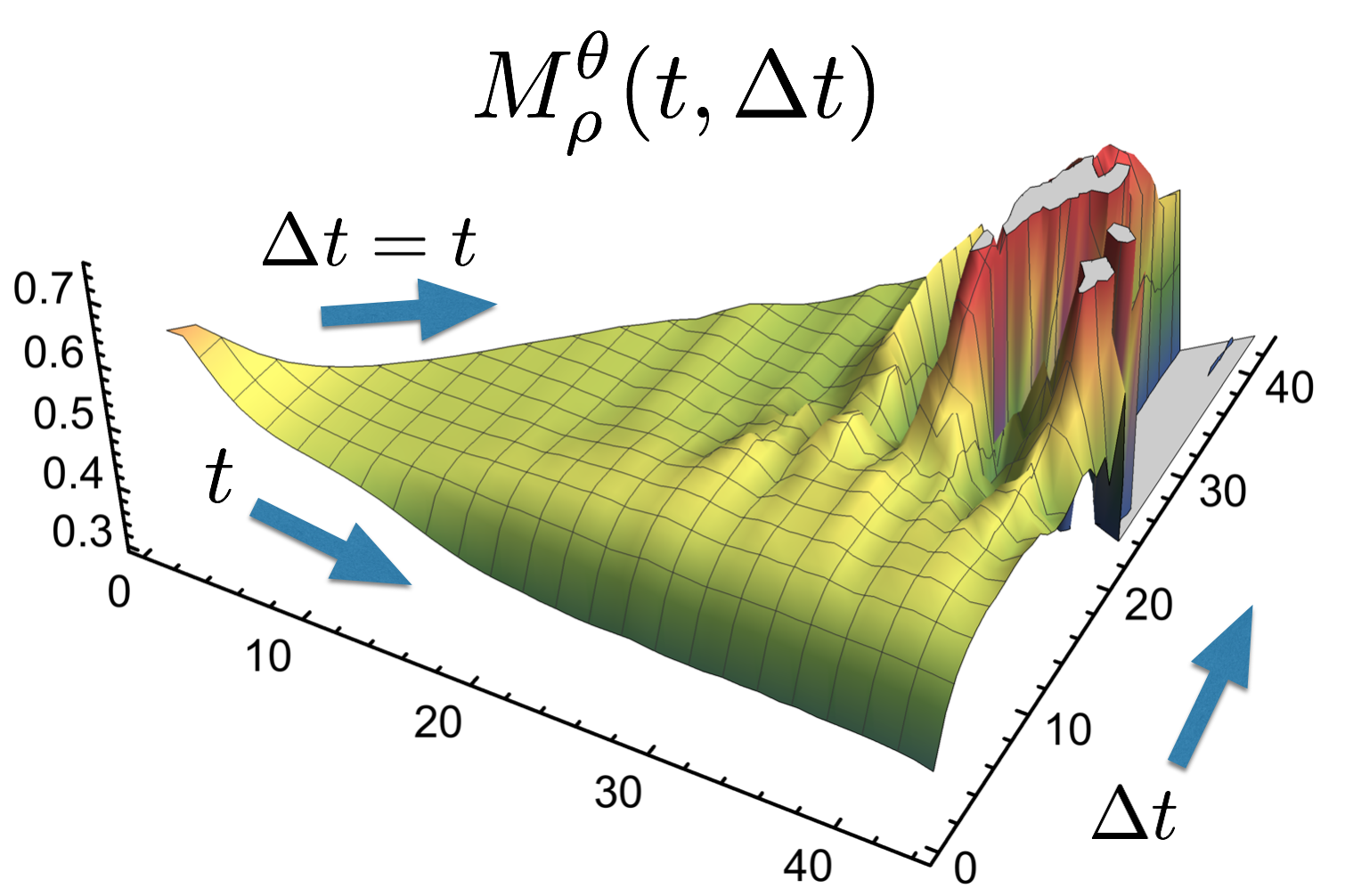}
		\caption{
	\label{fig:2DEMP} 
        The $\rho^+$ meson phase-reweighted effective mass for all $\Delta t \leq t$. 
        The standard effective mass in the upper panel of Fig.~\ref{fig:emps} corresponds to
        $M^\theta_\rho(t,t)$, a projection along the line $t = \Delta t$ indicated.
        The bottom panel of Fig.~\ref{fig:emps} 
        shows $M^\theta_\rho(t,\Delta t)$ on lines of constant $\Delta t$ parallel to the $t$ axis indicated. 
	}		
\end{figure}

\begin{figure}[!t]
	\includegraphics[width=0.95 \columnwidth]{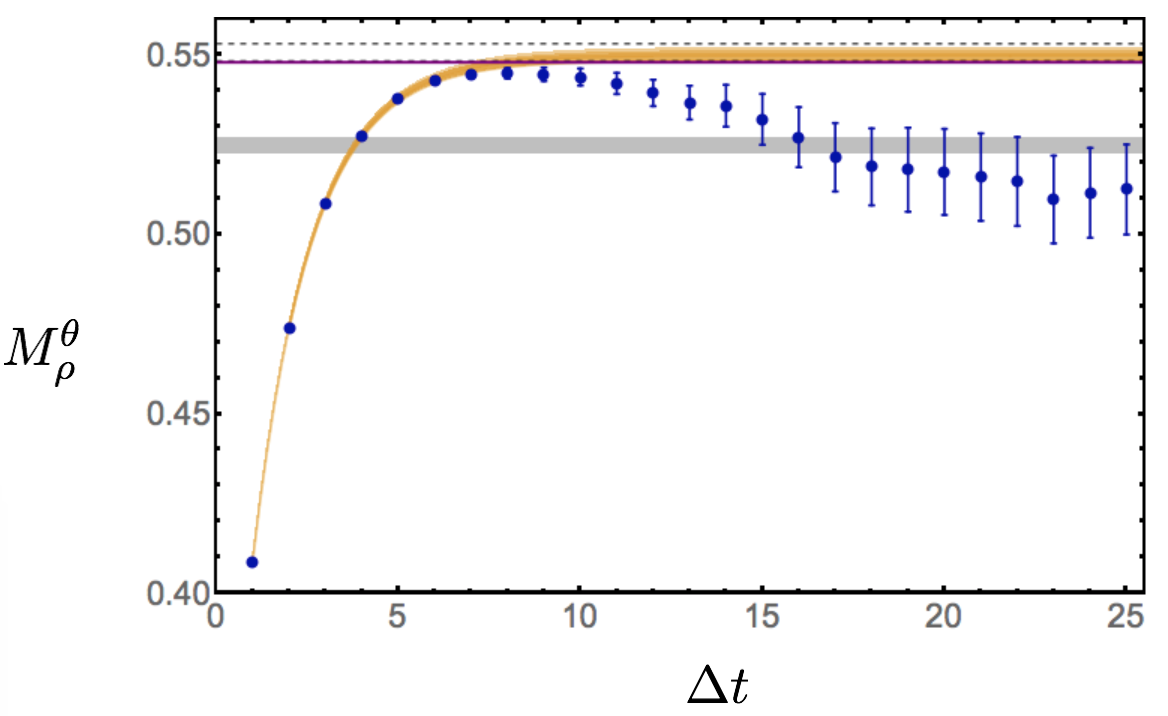}
		\caption{
	\label{fig:Rhoextrap} 
        The $\rho^+$ mass extracted from late-time phase-reweighted correlation functions.
	The light-brown shaded region corresponds to the $68\%$ confidence region associated with 	
        three-parameter (constant plus exponential)  fits to Eq.~\eqref{eq:PRWEMP}.
        The dashed lines show the extrapolated $M_\rho^\theta$ result
        including statistical and systematic uncertainties
        described in the main text.
        The gray horizontal band corresponds to a determination of the $\rho^+$ mass from the plateau region~\cite{Orginos:2015aya}.
        The purple line corresponds to the $\pi\pi$ non-interacting $p$-wave energy.	}		
\end{figure}

LQCD calculations of $M_\rho^\theta$ summarized in Figs.~\ref{fig:emps}-\ref{fig:Rhoextrap}
permit precise numerical study of small $\Delta t$ bias and $\Delta t \rightarrow t$ extrapolation.
These calculations employ $N\sim 130,000$ correlation functions
previously computed by the NPLQCD collaboration~ 
from smeared sources and point sinks on an 
ensemble of 2889 isotropic-clover gauge-field configurations 
at a pion mass of $M_\pi \sim 450~{\rm MeV}$
generated jointly by the College of William and Mary/JLab lattice group and by the NPLQCD collaboration, see Ref.~\cite{Orginos:2015aya} for further details.
The spacetime extent of the lattices is $48^3\times 96$ at a lattice spacing of $a\sim 0.117(1)~{\rm fm}$.
For all of the correlation functions examined in this work,  
momentum projected blocks are derived from quark propagators originating from smeared sources localized about a site in the lattice volume, 
as detailed in previous works by the NPLQCD collaboration, e.g. Ref.~\cite{Beane:2006mx,Orginos:2015aya}.
For instance, the blocks associated with the $\rho^+$ meson are
\begin{eqnarray}
{\cal B}^{(\rho^+)}_\mu ({\bf p},t; x_0) &  = & 
\sum_{\bf x} e^{i{\bf p}\cdot {\bf x}}\ \overline{S}_d({\bf x},t;x_0)\gamma_\mu S_u({\bf x},t;x_0).\;\;
  \label{eq:blockdef}
\end{eqnarray}
Correlations functions are derived by contracting the blocks with local interpolating fields~\cite{Detmold:2012eu},
e.g., 
\begin{eqnarray}
C^{(\rho^+;\mu)}({\bf p},t; x_0) & = & 
{\rm Tr}\left[\ 
{\cal B}^{(\rho^+)}_\mu ({\bf p},t; x_0)  \gamma^\mu
\ \right]
 ,
 \label{eq:blockcon}
\end{eqnarray}
where the trace is over color and spin. It is the phases of contracted momentum-projected blocks that have been used to form phase-reweighted correlation functions.
Expressions similar to those in eqs.~(\ref{eq:blockdef}) and (\ref{eq:blockcon}) are used for the nucleon and two-nucleon 
systems~\cite{Beane:2006mx,Orginos:2015aya}.

At large $t$ and small $\Delta t$, bias in $M_\rho^\theta$ is consistent with Eq.~\ref{eq:PRWEMP}.
At intermediate $\Delta t$, $M_\rho^\theta$ approaches a value consistent  with the $\pi\pi$ non-interacting $p$-wave energy $\sqrt{(2M_\pi)^2 + (2\pi/L)^2}$. 
At large $\Delta t$, $M_\rho^\theta$ approaches a lower-energy plateau consistent with the $\rho^+$ mass extracted from a $t=\Delta t$ plateau $t = 18 \rightarrow 28$.
The suppression of $\rho^+$ bound state contributions  compared to $\pi\pi$ scattering states contributions to $C_i(t)e^{-i\theta_i(t-\Delta t)}$ is found to be less severe in smaller volumes.
The energy gap between the bound and scattering states also increases in smaller volumes.
In accord with these arguments, the non-monotonic $\Delta t$ behavior visible in Fig.~\ref{fig:Rhoextrap} is not seen with $V=32^3$ or $V=24^3$.
$M_\theta^\rho$ is consistent with the $\rho^+$ mass determined in Ref.~\cite{Orginos:2015aya} for $\Delta t \gtrsim 5$ in these smaller volumes.
Variational methods employing phase reweighted correlation functions
with multiple interpolating operators
may be required 
to reliably distinguish closely spaced energy levels with large spatial volumes.

\begin{figure}[!t]
	\includegraphics[width=0.95 \columnwidth]{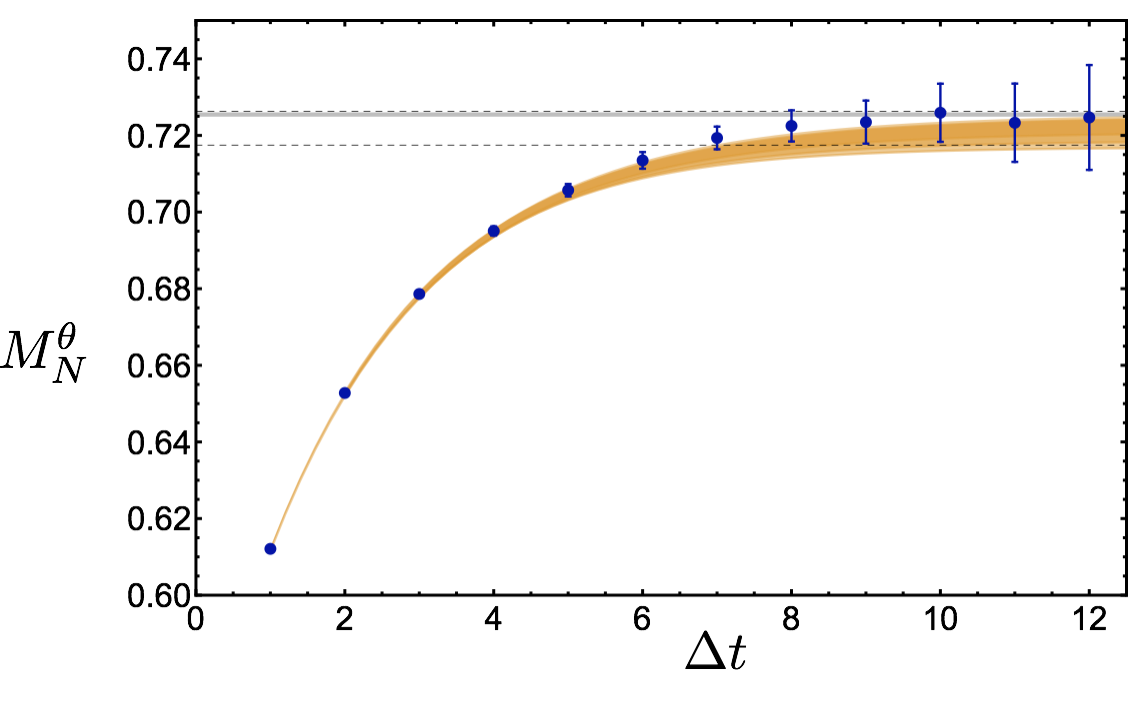}
		\caption{
	\label{fig:Nextrap} 
	The late-time nucleon phase-reweighted effective mass with statistical and systematic extrapolation errors shown
        with light-brown bands and dashed lines as in Fig.~\ref{fig:Rhoextrap}.
        The gray horizontal band corresponds to golden window result of Ref~\cite{Orginos:2015aya} obtained with four times higher statistics.
	}		
\end{figure}

The nucleon mass does not appear to have complications from low-lying excited states
and the late time phase-reweighted nucleon effective mass
derived from $\sim 100,000$ sources with $V=32^3$~\cite{Orginos:2015aya}
approaches its intermediate time plateau value at large $\Delta t$.
Small $\Delta t$ bias is well-described with a constant plus exponential form, and
the nucleon excited state gap can be 
extracted across a range of fitting regions as $\delta M_N = 786(44)(25)$~MeV, 
where the first uncertainty is statistical from a correlated $\chi^2$-minimization fit of $M_N^\theta(t,\Delta t)$ 
to Eq.~\eqref{eq:PRWEMP}
with $\Delta t = 2\rightarrow 10$ and $t = 30\rightarrow 40$  and the second uncertainty is a systematic  
determined from the variation in central value when the fitting region is changed to be $\Delta t = 1\rightarrow 10$ or $\Delta t = 3\rightarrow 10$. 
This result is consistent with a naive extrapolation $M_\sigma \sim 830$ MeV of the $\sigma$-meson mass 
determined at $M_\pi \sim 391$~MeV
~\cite{Briceno:2016mjc}. 
Results for strange-baryon excited-state masses from phase-reweighted effective mass extrapolations  are also 
consistent with the $\sigma$-meson mass in one- and two-baryon systems,
for instance $\delta M_{\Xi} = 822(44)(71)$~MeV and $\delta M_{\Xi\Xi(\si)} = 908(265)(82)$~MeV.

The $\Xi^-\Xi^-(\si)$ has slower StN degradation than a two-nucleon system
and is considered here for a first investigation of phase-reweighted baryon-baryon binding energies.
The $\Xi^-\Xi^-(\si)$ binding energy was
determined 
by
the NPLQCD collaboration
to be
$B_{\Xi\Xi (\si)}= 15.4(1.0)(1.4)~{\rm MeV}$
for the gauge field configurations considered here
using the correlation function production and sink-tuning~\cite{Beane:2009kya,Beane:2009gs,Beane:2010em} described for the deuteron and di-neutron in Ref.~\cite{Orginos:2015aya}.\footnote{$B_{\Xi\Xi(\si)} = -M_{\Xi\Xi(\si)}+2M_\Xi$
approaches the $\Xi\Xi(\si)$ binding energy in the infinite volume limit.
In finite volume $B_{\Xi\Xi(\si)}$ differs from the infinite-volume binding energy
by corrections that are exponentially suppressed by the binding momentum.
}
Results for $\Xi^-\Xi^-(\si)$ using the $\sim 100,000$ correlation function ensemble described above 
for constant 
fits to the phase reweighted binding energy with $t = 28\rightarrow 43$, $\Delta t = 1,2,3\rightarrow 6$ 
give 
$B_{\Xi\Xi(\si)}=15.8(3.5)(2.6)$ MeV.
Consistency between golden window results and phase-reweighted results with large $t$ and all $\Delta t \gtrsim 1$
suggests
a high degree of cancellation at all $\Delta t$ between excited state effects in 
one- and two-baryon phase reweighted effective masses.
$B_{\Xi\Xi(\si)}(t, \Delta t = 0) $, which only involves correlation function magnitudes,
plateaus to $7.1(0.6)(0.8)\text{ MeV}$.
Phase effects modify this magnitude result
by an amount on the order of nuclear energy scales rather than hadronic mass scales,
providing encouraging evidence that extrapolations involving modest $\Delta t$ can accurately determine nuclear binding energies in the noise region.
The precision of phase-reweighted results scales with the number of points in the noise region,
and could be increased on lattices of longer temporal extent then those used in this work ($\sim 11.2 {\rm fm}$).

\begin{figure}[!t]
	\includegraphics[width=0.95 \columnwidth]{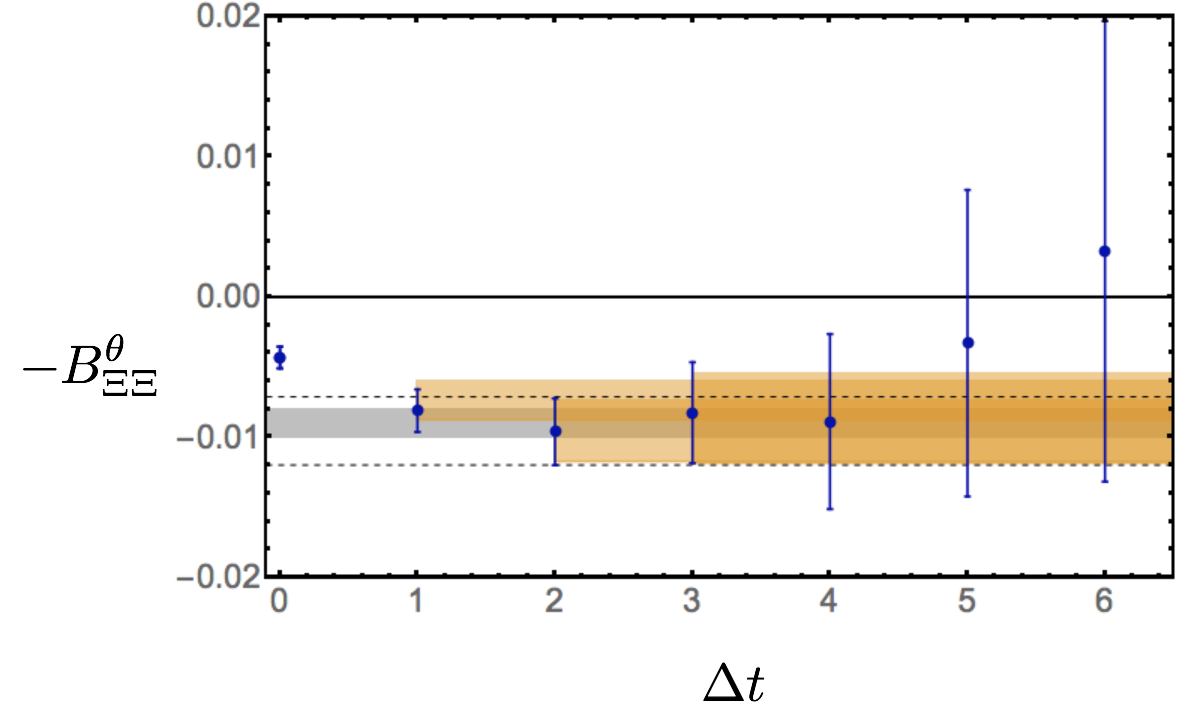}
		\caption{
	\label{fig:XiXiBindextrap} 
	The $\Xi^-\Xi^- (\si)$ phase-reweighted binding energy with statistical and systematic extrapolation errors shown
      with light-brown bands and dashed lines as in Fig.~\ref{fig:Rhoextrap}.
        The gray horizontal band corresponds to the golden window result of Ref.~\cite{Orginos:2015aya}, obtained with four times higher statistics.
	}		
\end{figure}

Phase reweighting 
allows 
energy levels
to be extracted
from LQCD correlation functions
at times later than the
golden window
accessible to standard techniques involving source and sink optimization~\cite{Beane:2009kya,Beane:2009gs,Beane:2010em,Detmold:2014rfa,Detmold:2014hla}.
It is expected that these methods
will permit the extraction of ground-state energies in systems
without a golden window.
The phase-reweighting method
is equivalent to a dynamical source improvement
in which the phase is held fixed while the magnitude of the hadronic correlation function
is evolved into its ground state,
and then the phase is released to provide a source for subsequent time slices.
The bias introduced by phase reweighting can be removed by extrapolation
but suffers from a StN problem
that can be viewed as arising from evolution of the dynamically improved source.
Generalizations of the phase-reweighting methods presented here
may allow for reaction rates,
operator matrix elements,
and other observables to be extracted from phase-reweighted correlation functions.

\vspace*{6mm}

\begin{acknowledgments}
\textbf{Acknowledgments.} We are grateful to the other members of the NPLQCD collaboration for allowing us to work with the 
high-statistics single- and multi-hadron correlation function ensembles integral to this work.
In particular, we thank Emmanuel Chang for efficient SQLite database management of large ensembles of 
unblocked correlation functions, Daniel Trewartha for enlightening visualizations of gauge field and correlation 
function fluctuations, and Silas Beane, Zohreh Davoudi, William Detmold, Phiala Shanahan, and Brian Tiburzi 
for helpful discussions and comments on this manuscript.
We also thank David Kaplan for many insights and helpful discussions on signal-to-noise and statistics, 
Leonardo Giusti, Tom DeGrand, and Dean Lee for very interesting discussions during 
{\it Sign 2017: International Workshop on the Sign Problem in QCD and Beyond}
 in Seattle (http://www.int.washington.edu/PROGRAMS/17-64w/), 
Joe Carlson, Steve Peiper, Bob Wiringa and Alessandro Lovato for discussions about phase pinning in GFMC calculations, 
and Tanmoy Bhattacharya, Ra{\'u}l Brice{\~n}o, Aleksey Cherman, Dorota Grobowska, Rajan Gupta, Natalie Klco, and Alessandro Roggero for helpful discussions 
at various stages of this work.
This research was supported in part by the National Science Foundation under grant number NSF PHY11-25915 
and we acknowledge the Kavli Institute for Theoretical Physics,
particularly the {\it Frontiers of Nuclear Physics} program (2016)
for hospitality during various stages of this work.
	Analyses of correlation functions were carried out on the 
	Hyak High Performance Computing and Data Ecosystem at the University of Washington, 
	supported, in part, by the U.S. National Science Foundation Major Research Instrumentation Award, Grant Number 0922770,
	and by the UW Student Technology Fee (STF). 
	Calculations were performed using computational resources provided
	by  NERSC (supported by U.S. Department of
	Energy Grant Number DE-AC02-05CH11231),
	and by the USQCD
	collaboration.  
	This research used resources of the Oak Ridge Leadership 
	Computing Facility at the Oak Ridge National Laboratory, which is supported 
	by the Office of Science of the U.S. Department of Energy under Contract 
	No. DE-AC05-00OR22725. 
	The PRACE Research Infrastructure resources Curie based in France at the Tr\`es Grand Centre de Calcul and MareNostrum-III based in 
	Spain at the Barcelona Supercomputing Center were also used.
	Our calculations, in part, used the {\tt chroma} software suite~\cite{Edwards:2004sx} produced under the auspices of USQCD's  DOE SciDAC project.  
	We are supported in part  by DOE grant No.~DE-FG02-00ER41132.  
	\end{acknowledgments}
%

\bibliography{bib_PRW.bib}

\appendix 

\newpage

\begin{center} 
{\bf Supplemental Material} 
\end{center} 

\begin{table}[!ht]
\begin{center}
\begin{minipage}[!ht]{16.5 cm}
\end{minipage}
\setlength{\tabcolsep}{1em}
\resizebox{\linewidth}{!}{%
\def\arraystretch{1.2}%
\begin{tabular}{|c| l | l | l |}
\hline
$\Delta t$    & \qquad $M^\theta_{\rho^+}$  & \qquad $M_N^\theta$          & \qquad $B_{\Xi\Xi(\si)}$     \\
\hline
\hline
$1$ & \quad 0.40872(21) &\quad 0.61209(50)&\quad  -0.0081(15)           \\
$2$ & \quad 0.47392(30) &\quad 0.65278(66)&\quad  -0.0096(24)           \\
$3$ & \quad 0.50841(40) &\quad 0.67861(88)&\quad   -0.0083(36)          \\
$4$ & \quad 0.52722(52) &\quad 0.6951(12)&\quad    -0.0089(62)         \\
$5$ & \quad 0.53774(67) &\quad 0.7057(16)&\quad   -0.003(11)          \\
$6$ & \quad 0.54284(84) &\quad 0.7135(22)&\quad    0.003(16)          \\
$7$ & \quad 0.5446(11) &\quad 0.7193(30)&\qquad\qquad       -      \\
$8$ & \quad 0.5449(15) &\quad 0.7225(41)&\qquad\qquad       -      \\
$9$ & \quad 0.5446(19) &\quad 0.7235(56)&\qquad\qquad       -      \\
$10$ & \quad 0.5439(23) &\quad 0.7259(76)&\qquad\qquad       -      \\
$11$ & \quad 0.5421(30) &\quad 0.723(10) & \qquad \qquad -      \\
$12$ & \quad 0.5395(37) &\quad 0.725(14) &  \qquad \qquad -      \\
$13$ & \quad 0.5368(47) &\qquad \qquad -  &  \qquad \qquad -      \\
$14$ & \quad 0.5359(58) &\qquad \qquad -  &  \qquad \qquad -      \\
$15$ & \quad 0.5321(71) &\qquad \qquad -  &  \qquad \qquad -      \\
$16$ & \quad 0.5271(83) &\qquad \qquad -  &  \qquad \qquad -      \\
$17$ & \quad 0.5215(95) &\qquad \qquad -  &  \qquad \qquad -      \\
$18$ & \quad 0.519(11) &\qquad \qquad -  &  \qquad \qquad -      \\
$19$ & \quad 0.518(12) &\qquad \qquad -  &  \qquad \qquad -      \\
$20$ & \quad 0.517(12) &\qquad \qquad -  &  \qquad \qquad -      \\
$21$ & \quad 0.516(12) &\qquad \qquad -  &  \qquad \qquad -      \\
$22$ & \quad 0.515(12) &\qquad \qquad -  &  \qquad \qquad -      \\
$23$ & \quad 0.510(12) &\qquad \qquad -  &  \qquad \qquad -      \\
$24$ & \quad 0.512(13) &\qquad \qquad -  &  \qquad \qquad -      \\
$25$ & \quad 0.513(13) &\qquad \qquad -  &  \qquad \qquad -      \\
\hline
PR Ground  & 0.5222(60)(27)   &  0.7220(33)(11) &    -0.0096(22)(11)         \\
PR Excited  & 0.5508(11)(7)   &  \qquad \qquad - &    \qquad \qquad -         \\
\hline 
GW Ground  &  0.5248(14)(15) & 0.72551(35)(26)&    -0.00909(59)(83)         \\
GW $\pi\pi$  &  0.547997(78)(14) & \qquad \qquad - &    \qquad \qquad -         \\
\hline 
\end{tabular}
}
\begin{minipage}[t]{8.5 cm}
\vskip 0.0cm
\noindent
  \caption{
Phase-reweighted (PR) effective masses of the $\rho^+$, nucleon and the effective 
energy difference between $\Xi\Xi(\si)$ and two $\Xi$'s
derived from  eq.~(\ref{eq:PRWdef}). 
The extrapolated PR ground values are taken from three-parameter constant plus exponential correlated $\chi^2$-minimization fits for
$M_N^\theta$ and one-parameter constant fits for $B_{\Xi\Xi(\si)}^\theta$
with statistical uncertainties for fits starting at $\Delta t = 2$ and systematic uncertainties
defined from variation of the $\Delta t$ fitting window as described in the main text.
PR data is taken from $t = 28 \rightarrow 43$ for the $\rho^+$ and  $\Xi\Xi(\si)$
and $t = 31 \rightarrow 40$ for the nucleon.
For the $\rho^+$, the region $\Delta t = 2 \rightarrow 10$ is used to constrain the first scattering state for the PR excited state result, 
while the region $\Delta t = 16\rightarrow 25$ is used to constrain the ground state.
Golden window (GW) ground refers to the ground-state energy determinations using the short and intermediate
time plateau regions described in Ref.~\cite{Orginos:2015aya}.
GW $\pi\pi$ refers to the non-interacting $p$-wave energy shift $\sqrt{(2M_\pi)^2 + (2\pi/L)^2}$ using $M_\pi$ and $L$ for the $48^3$ ensemble described in the main text.
}
\label{tab:masses}
\end{minipage}
\end{center}
\end{table}     
\end{document}